\documentclass[aps,twocolumn,showpacs,preprintnumbers,amsmath,amssymb,nofootinbib,superscriptaddress,showkeys,prl]{revtex4}

\usepackage{epsfig}
\usepackage{graphicx}

\begin{document}

\title{Alpha-alpha interaction with chiral two pion exchange and $^8$Be lifetime.}

\author{E. Ruiz Arriola} 
\email{earriola@ugr.es}

\affiliation{ Departamento de F\'{\i}sica At\'omica, Molecular y
Nuclear, Universidad de Granada, E-18071 Granada, Spain.}

\date{\today}

\begin{abstract} 
\rule{0ex}{3ex} Assuming that alpha particles are described by a
scalar-isoscalar chiral invariant heavy field it is shown how chiral
symmetry determines the alpha-alpha interaction at long distances
unambiguously from dimensional power counting of an effective chiral
Lagrangean.  The leading strong contribution is given by a two pion
exchange potential which turns out to be attractive and singular at
the origin, hence demanding renormalization. When $^8 {\rm Be}$ is
treated as a resonance state a model independent correlation between
the Q-factor and lifetime $1/\Gamma$ for the decay into two alpha
particles arises. For parameters compatible with potential model
analyses of low energy $\pi \alpha $ scattering it is found a
Breit-Wigner width $\Gamma = 4.3(3) {\rm eV}$ very close to the
experimental value, $\Gamma_{\rm exp.} = 5.57 (25) {\rm eV}$.
\end{abstract}

\pacs{03.65.Nk,11.10.Gh,21.30.Fe,21.45.+v,24.50.+g}

\keywords{Chiral Symmetry, Effective Field Theory, Alpha-Alpha
interaction, Two Pion Exchange, $^8$Be, Renormalization.}

\maketitle

Low energy nuclear reactions in the energy range of astrophysical
interest are generally extremely hard to measure experimentally in the
laboratory~\cite{Rolfs:1978sh}. This applies in particular to 
$\alpha\alpha$ scattering where $^8{\rm Be}$ is produced as a
narrow resonance~\cite{RevModPhys.41.247}. In this paper we approach
this problem from the theoretical side in the context of the chiral
symmetry\cite{Weinberg:1978kz,Ericson:1988gk} and Effective Field
Theories (EFT)~\cite{Bedaque:2002mn}. Specifically, it will be shown
how the lifetime of $^8 {\rm Be}$ in alpha-decay may be accounted for
in a model independent fashion by assuming an elementary field for the
$\alpha$-particle, and exploiting the spontaneous breakdown of chiral
symmetry of QCD as well as the fact that the $^8 {\rm Be}$ ground
state lies right above the $\alpha\alpha$ threshold.  In addition to
the standard Coulomb interaction, the scalar-isoscalar character of
the $\alpha$ particle implies that the long distance strong
interaction is dominated by two pion exchange (TPE) regardless on any
specific internal structure.

Pion exchange interactions between $\alpha$ particles have been
treated in the past in a variety of ways. A resonating group method
approach was used in Ref.~\cite{Shimodaya:1960} with an approximation
for the TPE in the mid-range. Forward dispersion relations for
$\alpha\alpha$ scattering have been
discussed~\cite{FangLandau:1974wh}. A folding model from a NN
potential was used in Ref.~\cite{Ericson:1981sa} and the $I$-wave
phase shift was computed in first order perturbation theory. An EFT
description of narrow resonances has been undertaken for pure contact
theories~\cite{Bertulani:2002sz}, i.e. without pions or Coulomb
forces.

The idea of associating elementary fields to nuclei at low energies is
rather natural~\cite{Locher:1978dk}.  The $\alpha$ particle is a
$^4{\rm He}$ nucleus with $(J^P, T)=(0^+,0)$, charge $Z_\alpha e = + 2
e $ and mass $M_\alpha = 2(M_p+M_n) - B = 3727.37 {\rm MeV} $, with
binding energy $B=28.2957 {\rm MeV}$. Because of the tensor force, the
wave function for the ground state can be a positive-parity mixture of
three $^1S_0$, six $^3P_0$, and five $^5D_0$ orthogonal
states~\cite{PhysRev.158.907}, the symmetric S-wave component being
the dominant part of the wave function, with significant D-wave and
almost negligible P-wave contributions. Thus, as a starting point we
associate a scalar-isoscalar Klein-Gordon charged field, $\alpha(x)$
to the $^4{\rm He}$ nucleus. Under charge conjugation the
$\alpha$-particle should transform into an anti-$\alpha$-particle $
\alpha (x) \to \bar \alpha ( x) $ meaning that the field is
non-hermitean. Further, under $SU(2)_R \otimes SU(2)_L $ chiral
transformations we assume $\alpha(x)$ to be invariant; candidates for
chiral partners with the same $B=4$ baryon number would be $^4{\rm H}$
and $^4{\rm Li}$ having both spin-2, $(J^P,T)=(2^-,1)$, thus belonging
to different Poincare group representations. 
The effective Lagrangean will include pions~\cite{Weinberg:1978kz} and
$\alpha$ particles which being much heavier, $M_\alpha \gg m_\pi$, are
better treated by transforming the Klein-Gordon field as $\alpha (x) =
e^{-i M_\alpha v \cdot x} \alpha_v (x) $ with $\alpha_v (x) $ the
heavy field and $v^\mu$ a four-vector fulfilling $v^2=1$, eliminating
the heavy mass term~\cite{Jenkins:1991es,Bernard:1995dp}. Keeping the
leading $M_\alpha$ term, the
effective Lagrangean reads
\begin{eqnarray}
{\cal L} &=& i M_\alpha \bar \alpha_v v \cdot \partial \alpha_v +
\frac{f^2}{4} \left[ \langle \partial^\mu U^\dagger \partial_\mu U
\rangle + \langle \chi U^\dagger + \chi^\dagger U \rangle \right]
\nonumber \\ &+& g_0 \bar \alpha_v \alpha_v \langle \partial^\mu
U^\dagger \partial_\mu U \rangle + g_1 \bar \alpha_v \alpha_v \langle
\chi U^\dagger + \chi^\dagger U \rangle \nonumber \\ &+& g_2 \bar
\alpha_v \alpha_v \langle v \cdot \partial U^\dagger v \cdot \partial
U \rangle + \lambda \left( \bar \alpha_v \alpha_v \right)^2
\label{eq:L_I} 
\end{eqnarray} 
where the pion field in the non-linear representation is written as a
$SU(2)$-matrix, $U= e^{ i \vec \tau \cdot \vec \pi / f} $, with $\vec
\tau$ the isospin Pauli matrices, $f$ the pion weak decay constant in
the chiral limit $f = 88 {\rm MeV}$, $\chi= m^2 /2 $ and $\langle ,
\rangle $ means trace in isospin space.  Here $g_0$, $g_1$, $g_2 $ and
$\lambda$ are dimensionless coupling constants which are not fixed by
chiral symmetry. This Lagrangean is the analog of the
Weinberg-Tomozawa Lagrangean and EFT extensions for $\pi N $
interactions~\cite{Ericson:1988gk,Bernard:1995dp} to the case of the
$\pi \alpha $ system.  Photons are included by standard minimal
coupling $\partial^\mu \alpha \to D^\mu \alpha = \partial^\mu \alpha +
Z_\alpha e i A^\mu \alpha $. For definiteness we take, $M_\alpha =
3727.3 {\rm MeV}$, $f \to f_\pi =92.4 {\rm MeV}$ and $m \to m_\pi=138
{\rm MeV}$.

\medskip
\begin{figure}[tbc]
\begin{center}
\epsfig{figure=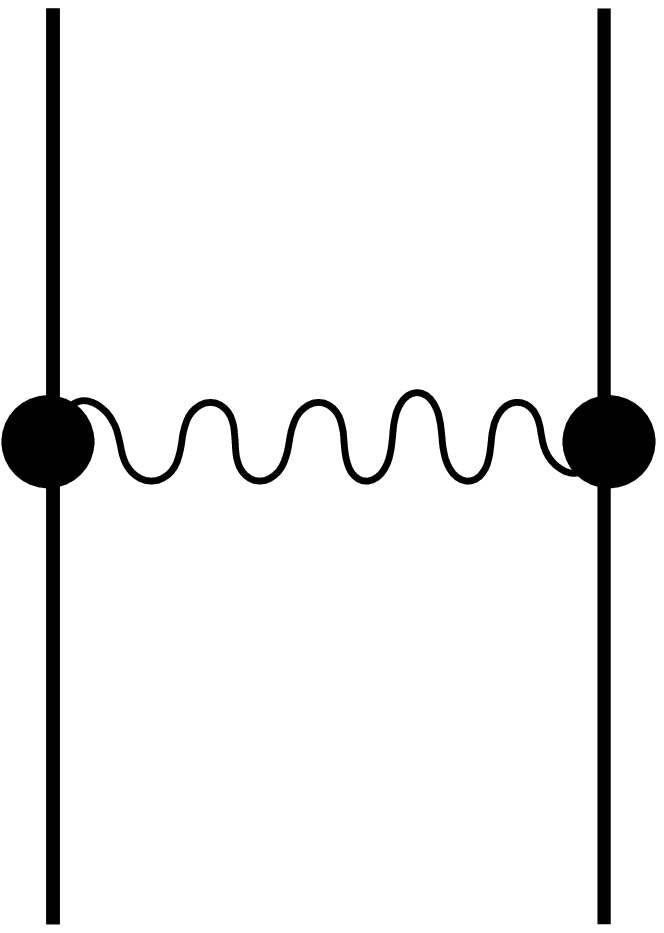,height=4cm,width=2.5cm}\hskip1cm 
\epsfig{figure=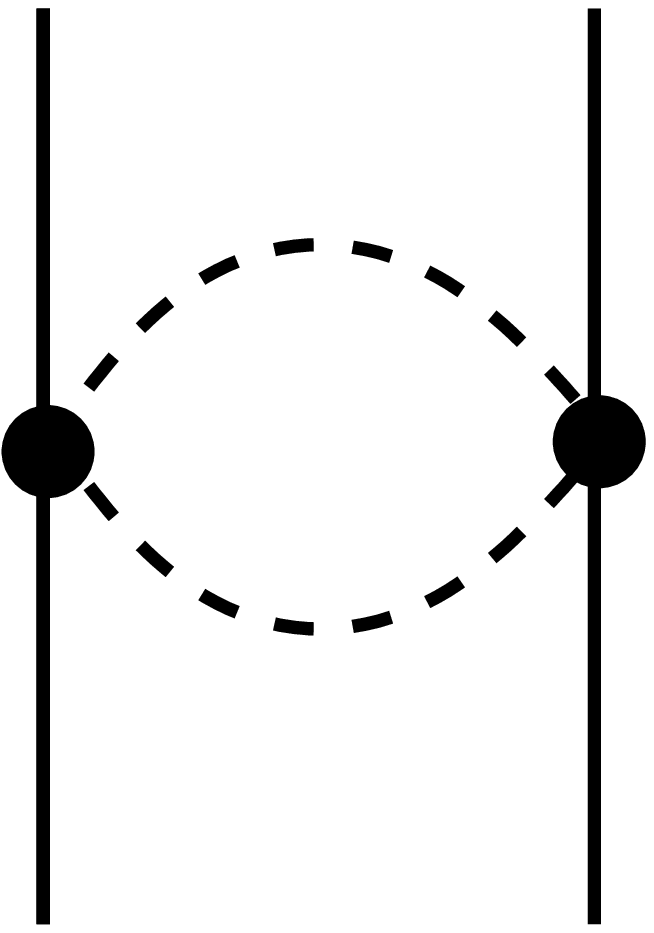,height=4cm,width=2.5cm}
\end{center}
\caption{Diagrams contributing to the $\alpha-\alpha$ potential at
 long distances: One-Photon Exchange (left) and Two-Pion Exchange loop
 (right). Full lines represent the $\alpha$ particle field, the dashed
 lines are pion fields, wiggly lines are photon fields. The full blobs
 are the $\alpha \alpha \gamma $ and the $\alpha \alpha \pi \pi $
 vertices respectively.}
\label{fig:tpe}
\end{figure}

To estimate the couplings $g_0$, $g_1$ and $g_2$ we look first at low
energy $\pi \alpha$ scattering ~(see~\cite{Ericson:1988gk} for a
review). From Eq.~(\ref{eq:L_I}) we get the $\pi^a (p_\pi) + \alpha
\to \pi^b (p_\pi') + \alpha $ invariant amplitude at lowest tree level
order
\begin{eqnarray}
i {\cal F}_{\pi \alpha \to \pi \alpha}^{ab} = \frac{4 i \delta^{ab}}{f_\pi^2}
\left( g_0 p'_\pi \cdot p_\pi - g_1 m_\pi^2 + g_2 v \cdot
p_\pi v \cdot p_\pi' \right) \, ,
\end{eqnarray} 
where $a$ and $b$ are the final and initial pion isospin states
respectively.  In the limit $M_\alpha \gg m_\pi $ LAB and CM coincide,
so taking $v^\mu = (1,\vec 0)$ and the kinematics as $p_\pi = (
\omega, \vec p) $ and $ p_\pi' = (\omega , \vec p') $ with
$\omega=\sqrt{\vec p^2+m^2}$ the scattering amplitude is given by
(${\cal F} = 8 \pi \sqrt{s} f$)
\begin{eqnarray}
f_{\pi \alpha} (p, \theta) &=& \frac{ g_0 ( m_\pi^2 + p^2
(1-\cos\theta )) - g_1 m_\pi^2   + g_2 \omega^2}
{2 \pi M_\alpha f_\pi^2}
 \nonumber \\
&=& A_0 + B_0 p^2 + 3 A_1 p^2 \cos \theta + \dots 
\label{eq:f_low}
\end{eqnarray} 
where in the second line the threshold parameters $A_0$, $B_0$ and
$A_1$ have been introduced. From  mesic $\pi^--^4{\rm He}   $ atoms one has
$A_0 = (-0.138 + i 0.045) {\rm
fm}$~\cite{Backenstoss:1970wv,Ericson:1988gk} while $B_0 = -0.18 {\rm
fm}^3$ and $A_1 = (0.42 + i0.06) {\rm fm}^3$ from forward dispersion
relations~\cite{Pilkuhn:1975gk}. Our description is not realistic
concerning the direct comparison with data; Coulomb distorsion has
been disregarded and treating the $\alpha$ particle as elementary
precludes pion absorption since real $g_{0,1,2}$ imply ${\rm Im} A_l
=0$. However, realistic calculations allow to switch off these
effects~\cite{Khankhasaev:1989xf} yielding the pure potential values,
$A_0^{\rm pot}= -0.091(17) {\rm fm}$ and $A_1^{\rm pot}= 1.058 (144)
{\rm fm}^3$. We take $B_0^{\rm pot}=-0.2(2) {\rm fm}^3$
from~\cite{Pilkuhn:1975gk}, the error being an educated guess, since
there are unresolved discrepancies~(see
e.g. \cite{Brinkmoeller:1993si}). The small scattering length supports
our perturbative calculation and in fact $A_0 \to 0 $ in the chiral
limit, $m_\pi \to 0$.  From the values above we get $g_0 = -82(11)$,
$g_1=-5.3(3)$ and $g_2=77(12)$.  On the other hand, the double
scattering contribution to the $\pi \alpha$ s-wave~\cite{Lohs:1978qp},
yields the identification $ g_1-g_0-g_2= M_\alpha \langle r^{-1}
\rangle_\alpha /f^2 $ which has the correct pion mass dependence,
provided all quantities are evaluated in the chiral limit, and
suggests that $g_1-g_0 -g_2 > 0$. For $\langle r^{-1} \rangle_\alpha
=0.5 {\rm fm}$, a realistic value, and using either $f$ or $f_\pi$
yields $g_1-g_0-g_2\sim 40-47$, while we get $\sim 0-20$ instead.  A
expected decrease of $\langle r^{-1} \rangle_\alpha $ in the chiral
limit might naturally accomodate the discrepancy. On top of this, pion
loop corrections to $\pi \alpha $ scattering, which are ${\cal O}
(1/f_\pi^4)$, might have a sizable impact on $g_0$,$g_1$ and $g_2$ due
to chiral logs. Clearly, a more systematic assessment of the input
values of the couplings and their uncertainties would be most useful.

Let us now turn to the calculation of the long distance $\alpha \alpha
$ potential.  The leading direct t-channel TPE contribution is
depicted in Fig.~\ref{fig:tpe} and can be written as
\begin{eqnarray}
{\cal F}_{\alpha \alpha \to \alpha \alpha} (q) = \frac{i}2 \int \frac{d^4
p}{(2 \pi)^4} \frac{| {\cal F}_{\pi \alpha \to \pi \alpha}^{ab} |^2}
{((p-q)^2-m_\pi^2 )(p^2-m_\pi^2)} \, . 
\end{eqnarray} 
There is a factor $\delta^{ab} \delta_{ab}=3$ coming from $\pi^+
\pi^-$ as well as $\pi^0 \pi^0$ exchange (we neglect here
tiny isospin breaking effects).
As we see by power counting the integral is quartically divergent.
Using the dispersion
relation
\begin{eqnarray}
{\cal F} (q^2) = \int_{4 m_\pi^2}^\infty d \mu^2
\frac{\rho(\mu^2)}{q^2 - \mu^2 + {\rm i} 0^+} + {\rm c.t.} \, , 
\end{eqnarray} 
with $\rho(\mu^2)$ the spectral density and c.t. stands for
counterterms, e.g. $\lambda$ in Eq.~(\ref{eq:L_I}), which will not
contribute to the coordinate space potential at positive but
non-vanishing distances. From Cutkosky's rules one gets
\begin{eqnarray}
\rho(\mu^2) = \frac{1}{2 \pi^2 f_\pi^4} \left( A_-^2 + 2 A_+^2
\right) \left[1- \frac{4 m_\pi^2}{\mu^2}\right]^\frac12 \, , 
\end{eqnarray}
where $A_\pm = (g_1- g_0) m^2 + g_0 \mu^2 /2 \pm g_2 (\mu^2 /4 -m^2
)$. Note that $\rho(\mu^2)$ is a positive quantity. This result agrees
with Ref.~\cite{Ericson:1981sa} only when $g_2=0$ and resembles a
similar calculation for the central NN force~\cite{Kaiser:2001pc}. In
the heavy mass limit $\sqrt{s} \to 2 M_\alpha$ and in the CM system $
v \cdot q =0$ so that the potential is given by the expression
\begin{eqnarray}
V_{\alpha \alpha}^{2\pi} (\vec x ) = - \frac{1}{4 M_\alpha^2} \int
\frac{d^3 q }{(2 \pi)^3} {\cal F} (- \vec q^2) e^{i \vec x \cdot \vec
q} \, .
\end{eqnarray} 
Computing the $q$ and the $\mu$ integrals, up to distributions located
at $\vec x=0$, the final result becomes
\begin{eqnarray}
V_{\alpha \alpha}^{2\pi} (r)&=&-\frac{3 m_\pi^7 \left[ K_0 ( 2 x) f(x) + K_1(2 x) g(x)  \right] }{ 32 \pi^3
M_\alpha^2 f_\pi^4 x^6}  
\label{eq:tpe-pot}
\end{eqnarray}
where $x= m_\pi r$, $K_0 (x)$ and $K_1(x)$ are modified Bessel
functions and $ f(x) = 4(g_0+g_1)^2 x^4 + 10( 12 g_0^2 + 4g_2 g_0 + 3
g_2^2) + \left[84 g_0^2 + 24 (g_1+g_2) g_0 + g_2 (4 g_1 +15 g_2)
\right] x^2 $, $ g(x) = 4 (g_0 + g_1)( 6 g_0 + g_2) x^3 + 10(12 g_0^2
+ 4 g_0 g_2 + 3 g_2^2)x $. The TPE potential is attractive everywhere
as can be recognized from the positivity of the spectral function. The
potential (\ref{eq:tpe-pot}) is the strong interaction analog of the
time-honoured Casimir-Polder electromagnetic
forces~\cite{PhysRev.73.360,Feinberg:1989ps}. In fact, the TPE
potential becomes singular at short distances,
\begin{eqnarray}
V_{\alpha \alpha}^{2\pi} (\vec x ) = - \frac{15 ( 12 g_0^2 + 4 g_0 g_2
+ 3 g_2^2)}{32 \pi^3 M_\alpha^2 f_\pi^4 } \, \frac{1}{r^7} + \dots
\label{eq:v-short}
\end{eqnarray} 
This is a relativistic and attractive Van der Waals interaction which
is explicitly independent on the pion mass. 
In the opposite limit of long distances we have
\begin{eqnarray}
V_{\alpha \alpha}^{2\pi} (\vec x ) \to - \frac{3 (g_0+g_1)^2
m_\pi^{9/2}}{16 \pi^{5/2} M_\alpha^2 f_\pi^4 } \frac{e^{-2 m_\pi 
r}}{r^{5/2}} \, . 
\end{eqnarray} 
Of course, the previous potential describes the strong interaction
piece, and as we see is ${\cal O} (1/f_\pi^4)$. Electromagnetic
effects correspond to minimally couple photons.  We keep one photon
exchange (see Fig.~\ref{fig:tpe}) but neglect two or higher photon
exchanges as well as terms $ {\cal O} ( Z_\alpha^2 e^2 /f^4 )$ which
may be systematically computed from higher dimensional corrections to
the Lagrangean (\ref{eq:L_I}). These and other effects will be
analyzed in more detail elsewhere.

The total potential in a long distance expansion is given by adding
the TPE and Coulomb potentials  
\begin{eqnarray}
V(\vec x) = V_{\alpha \alpha}^{2\pi} (\vec x) + \frac{e^2
Z_\alpha^2}{r}+ \dots
\label{eq:pot-long}
\end{eqnarray}
with $Z_\alpha =2 $ and $e^2 = 1/137.04 $ the fine structure constant.
The dots in Eq.~(\ref{eq:pot-long}) represent shorter range
corrections than TPE.  Rotational invariance allows to write the
relative s-wave function as $\Psi(\vec x) = u_{0,p} (r) /\sqrt{4\pi} $
with $u_{0,p} (r)$ the reduced s-wave function fulfilling
\begin{eqnarray}
-u_{0,p}''(r) &+& \left[ U_{\alpha \alpha}^{2\pi} (r) + \frac{2}{a_B
 r} \right] u_{0,p}(r) = p^2 u_{0,p} (r) \, , 
\label{eq:sch-l}
\end{eqnarray} 
with $U_{\alpha \alpha}^{2\pi} (r) = M_\alpha V_{\alpha \alpha}^{2\pi}
(r) $, $a_B = 2/ (M_\alpha Z_\alpha^2 e^2) = 3.63 {\rm fm}$ the Bohr
radius and $p = \sqrt{ M_\alpha E} $ the CM momentum. The problem is
to solve Eq.~(\ref{eq:sch-l}) with suitable boundary conditions, but
since the potential is singular and attractive at the origin, see
Eq.~(\ref{eq:v-short}), some renormalization proves
necessary. Actually, the regularity condition, $u_{0,p}(0)=0$, only fixes the solution up to an arbitrary
constant~\cite{Case:1950,Frank:1971xx} which must be fixed {\it
independently} of the potential. Furthermore, from the self-adjoint
condition(for a discussion within the NN context
see~\cite{Valderrama:2005wv}) there is the relation
\begin{eqnarray}
0=\left[ {u_{0,k}}^* u_{0,p}'- {u_{0,k}'}^* u_{0,p} \right]\Big|_{r_c}
\label{log:der}
\end{eqnarray} 
where the limit $r_c \to 0 $ for the short distance cut-off is
understood. Taking $p=k$ and $k=0$ we get further 
\begin{eqnarray}
{\rm Re} \left[ \frac{u_{0,p}'(r_c)}{u_{0,p}
(r_c)}\right] = \frac{u_{0,p}'(r_c)}{u_{0,p}
(r_c)} = \frac{u_{0,0}'(r_c)}{u_{0,0}
(r_c)} \, . 
\label{eq:log-corr} 
\end{eqnarray} 
This means that the logarithmic derivative at the origin is energy
independent and real leaving only one free parameter left which
will be fixed below. In essence, this is the non-perturbative
renormalization program with one counterterm described
in~\cite{Valderrama:2005wv} for singular potentials.

\begin{figure}[tbc]
\begin{center}
\epsfig{figure=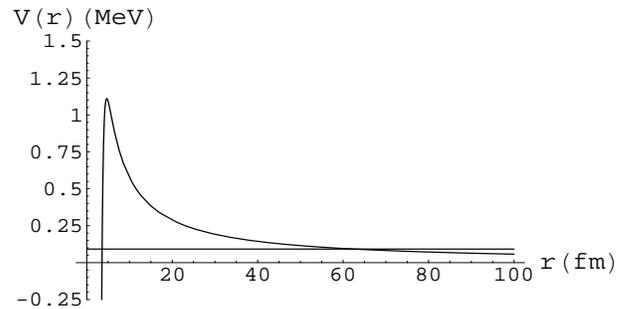,height=4cm,width=8cm}
\end{center}
\caption{The $l=0$ total two-pion exchange plus Coulomb potential barrier
for the $\alpha-\alpha$ system (in MeV) as a function of the relative
distance (in fm). The horizontal straight line represents the energy
corresponding to the energy $Q=91.84 {\rm KeV}$ of $^8$Be nucleus in
its ground state, $J^P = 0^+$.}
\label{fig:barrier}
\end{figure}

The $^8 {\rm Be}$ nucleus in its $ (J^P, T)= (0^+,0)$ ground state is
unstable against $\alpha$-decay, $^8 {\rm Be} \to {}^4{\rm He} +
{}^4{\rm He} $ with $Q=91.84 \pm 0.04 {\rm KeV}$ and a very small
(Breit-Wigner) width $\Gamma_{BW}=5.57 \pm 0.25 {\rm eV}$~(see
\cite{Tilley:2004} for a review). The relative CM momentum is $p =
\sqrt{ 2 \mu_{\alpha \alpha} Q } = 19.2 {\rm MeV}$ ($2 \mu_{\alpha
\alpha}=M_\alpha)$. The corresponding de Broglie wavelength, $\lambda
\sim 10 {\rm fm}$, is much larger than the size of the $\alpha$
particle, so one would not expect internal structure playing a
crucial role. The outer classical turning point is determined
by the Coulomb potential yielding $r_{\rm max}=62.6 {\rm fm}$.  The
situation is illustrated in Fig.~\ref{fig:barrier} where the total
$l=0$ TPE plus the Coulomb potential barrier for the $\alpha-\alpha$
system are depicted, together with the experimental Q-value $Q=91.84
{\rm KeV}$ for the $^8 {\rm Be} \to {}^4{\rm He} + {}^4{\rm He} $
reaction which proceeds by standard tunnel effect.

\medskip

It is instructive to analyze the decay process within the WKB approximation
where the width is given by~\cite{Gurvitz:1986uv}
\begin{eqnarray}
\Gamma_{\rm WKB} = \frac{e^{-2\, \int_{r_{\rm min}}^{r_{\rm max}} dr
|p(r)| }}{ 4 \mu_{\alpha \alpha} \int_{r_c}^{r_{\rm min}} \frac{dr}{2 p (r)}}\, , 
\label{eq:gamma_WKB} 
\end{eqnarray} 
with $p(r) = \sqrt{2 \mu_{\alpha\alpha}
(Q-V(r))}$. Here, $r_{\rm min} $ and $r_{\rm max}$ are the classical
turning points fulfilling $ V(r_{\rm min}) = V(r_{\rm max})=Q = M_{^8
Be} - 2 M_\alpha $. We obtain $\Gamma_{\rm WKB} = 8.6 (4) {\rm eV}$ for the
experimental $Q$, a correct order of magnitude compatible with the
expected accuracy of the WKB formula. Although the TPE potential
diverges at short distances, the inner classical turning point takes
typically the value $r_{\rm rmin}=3 {\rm fm}$ for which there is
about $1 {\rm MeV}$ cancellation between TPE and Coulomb potentials.
So, the tunneling region is not determined by the
singularity. Finite cut-off corrections to $\Gamma_{\rm WKB} $ are
${\cal O} (r_c^{9/2}) $ for $r_c \ll r_{\rm min}$ as can be seen from
Eq.~(\ref{eq:gamma_WKB}).

A rigurous treatment of $^8 {\rm Be} $ as a exponentially
time-decaying state requires finding a pole of the S-matrix in the
second Riemann sheet of the complex energy plane, so we look for exact
numerical solutions of Eq.~(\ref{eq:sch-l}) fulfilling the asymptotic
boundary condition of a spherically outgoing Coulomb wave,
\begin{eqnarray}
u_{0,p} (r) \to G_0 (\eta, \rho) + i F_0 (\eta, \rho) \, , 
\end{eqnarray} 
with $\eta= 1/(pa_B) $ and $\rho = p r $.  For complex momenta $ p=
p_R + {\rm i} p_I $ the energy also becomes complex $ E = Q - i\Gamma
/2 $. The boundary condition, Eq.~(\ref{eq:log-corr}), implementing
self-adjointness provides a correlation between $\Gamma$ and $Q$
through the TPE potential. We get $ \Gamma_{\rm pole} =
3.4(2) \, {\rm eV} $ for the S-matrix pole width, fairly
independently of the cut-off radius for $r_c \ll r_{\rm min} \sim 3 {\rm
fm}$.

The experimentally
determined~\cite{Tilley:2004} Breit-Wigner small width involves the
s-wave phase shift~\cite{Rasche:1967,Kermode:1967} $\Gamma_{\rm BW} =
2 / \delta'_0 (E_R)$ for $\delta_0 ( E_R ) =\pi /2$. We get 
\begin{eqnarray}
\Gamma_{\rm BW} ( ^8 {\rm Be} \to \alpha \alpha)= 4.3(3) \, {\rm eV}
\, , \, (\, {\rm exp.} 5.57 (25) {\rm eV}\, ) \, ,
\end{eqnarray} 
for $E_R =91.8 {\rm KeV}$ and the couplings $g_0$, $g_1$ and $g_2$
with their uncertainties obtained from low energy $\pi \alpha$
scattering, Eq.~(\ref{eq:f_low}) and
\cite{Khankhasaev:1989xf}. Further, we analyze the scattering length
$\alpha_0$ and the effective range $r_0$ defined from the s-wave phase
shift low energy expansion
\begin{eqnarray}
\frac{ 2 \pi \cot \delta_0 (p)}{a_B(e^{2\pi \eta}-1)} + \frac2{a_B} h(\eta) = -
\frac1{\alpha_0} + \frac12 r_0 p^2 + \dots
\end{eqnarray} 
with $ h(x) $ the Landau-Smorodinsky 
function~\cite{Rasche:1967,Kermode:1967}.  From the universal low
energy theorem of Ref.~\cite{Valderrama:2005wv} in the Coulomb case we
obtain (in fm)
\begin{eqnarray}
r_0 = 1.03(1) - \frac{5.3(3)}{\alpha_0} + \frac{29(4)}{\alpha_0^2} \, .  
\end{eqnarray} 
The numerical coefficients depend on the total long distance potential
(\ref{eq:pot-long}) only. Using Eq.~(\ref{eq:log-corr}) we find
$\alpha_0^{\rm th} = -1210(70) {\rm fm} $ and $r_0^{\rm th} = 1.03(1)
{\rm fm}$, in reasonable agreement with $\alpha_0= -1630(150) {\rm fm}
$ and $r_0 =1.08(1) {\rm fm}$ from a low energy analysis of the
data~\cite{Rasche:1967}. The correlation between $(Q,\Gamma)$ and
$(\alpha_0,r_0)$ was pointed out long
ago~\cite{Kermode:1967,Kermode:1969} but has no predictive power.  As
we see, this is compatible with the underlying chiral TPE potential
which, in addition, correlates $r_0$ with $\alpha_0$ and $\Gamma$ with
$Q$. At higher energies further ingredients are needed; for $E_{\rm
LAB} = 3 {\rm MeV}$, i.e.  $1/p \sim 2.5 {\rm fm}$, we get $\delta_0 =
141 \pm 2^0$ whereas $\delta_0=128.4 \pm 1^0$ from the scattering
data~\cite{RevModPhys.41.247}. Note that we have made no attempt to
refit the couplings $g_0$, $g_1$ and $g_2$ stemming from low energy
$\pi \alpha$ scattering, although it is obvious that further
attraction is needed (a $20-30\%$ increase in $g_0$ would do). The
fact that we produce numbers in the right order of magnitude is
encouraging and calls for further improvements.

A more systematic determination of the input parameters would be
highly desirable, possibly including pion loops in $\pi \alpha$
scattering. Moreover, an accurate description of $\alpha\alpha$
scattering data in the higher energy elastic region below $p + ^7{\rm
Li}$ and $n +^7{\rm B}$ production threshold or $\alpha$-decays of the
excited $^8{\rm Be}$ states becomes sensitive to deeper interaction
regions. This may require some peripheral nuclear structure
information involving e.g. $ \alpha \to t +p $ and $\alpha \to ^3 {\rm
He} + n$ intermediate state contributions on top of the chiral
interactions deduced here. With these provisos in mind, the present
work provides an example on how chiral symmetry might be useful in low
energy nuclear reactions eventually providing theoretical constraints
in the hardly accessible regions of astrophysical interest.

I thank M. Pav\'on Valderrama for encouragement, J. Nieves and A. Gal
for remarks on the paper and N. Kaiser for pointing out a 
mistake. Support by the Spanish DGI and FEDER funds grant
no. FIS2005-00810, Junta de Andaluc{\'\i}a grants no. FQM225-05 and EU
Integrated Infrastructure Initiative Hadron Physics Project contract
no. RII3-CT-2004-506078 is also acknowledged.


\end{document}